\documentclass[aps,prl,twocolumn,superscriptaddress,showpacs]{revtex4-1}
\usepackage{amsmath}
\usepackage[dvips,xdvi]{graphicx}
\usepackage{amssymb}
\usepackage{epsfig}

\begin{document}
\title{Manipulation of atom number distributions in lattice-confined spinor gases}
\author{J. O. Austin}
\author{Z. N. Shaw}
\author{Z. Chen}
\affiliation{Department of Physics, Oklahoma State University, Stillwater, Oklahoma 74078, USA}
\author{K. W. Mahmud}
\email{Present address: Quidient LLC, Columbia, MD 21046} \affiliation{Joint Quantum Institute, University of Maryland,
College Park, Maryland 20742, USA}
\author{Y. Liu}
\email{Electronic address: yingmei.liu@okstate.edu} \affiliation{Department of Physics, Oklahoma State University,
Stillwater, Oklahoma 74078, USA}
\date{\today}

\begin{abstract}
We present an experimental study demonstrating the manipulation of atom number distributions of spinor gases after
nonequilibrium quantum quenches across superfluid to Mott insulator phase transitions in cubic optical lattices. Our data
indicate that atom distributions in individual Mott lobes can be tuned by properly designing quantum quench sequences,
which suggest methods of maximizing the fraction of atoms in Mott lobes of even occupation numbers and have applications
in attaining different quantum magnetic phases including massively entangled states. In addition, we find qualitative
agreements between our experimental data and numerical simulations based on time-dependent Gutzwiller approximations in
two-dimensional systems.
\end{abstract}

\maketitle

Possessing a spin degree of freedom, spinor Bose-Einstein condensates (BECs) can be combined with optical lattices and
microwave dressing fields to offer a system with an exceptional level of control over many parameters, such as the number
of interacting atoms, temperature, total spin of the system, the lattice potential, and dimensionality of the
system~\cite{Mahmud,Zhao,zhaoUwave,Zhao2,Phuc,Jiang,Austin,Chen,Imambekov,Lacki,Bloch}. Lattice-confined spinor BECs have
thus been utilized to form highly programmable quantum simulators, which are capable of studying a vast array of topics at
the forefront of physics research, especially those that are too computationally complex to study using classical
computers~\cite{Austin,Zhao2,Bloch}. One feature of lattice-confined spinor gases of particular interest is that atoms in
Mott-insulator (MI) lobes of different occupation numbers can have distinct and potentially desirable properties, e.g.,
antiferromagnetic spinor gases can only form spin singlets in even Mott
lobes~\cite{Jiang,Demler,Imambekov,Lacki,Snoek,Yip}. Many-body spin singlet states, which consist of massively entangled
spin components, have been suggested as exemplary platforms for studying quantum memory and quantum
metrology~\cite{Zhao2}. The number distribution of ultra-cold atoms in lattices has thus been a topic of great
interest~\cite{Chen,Landig,Campbell,Tuchman,Zhou,Will,Agarwala,Tiesinga,Bakr,Sansone}. Various experimental approaches
have been realized to determine number distributions after atoms are loaded into deep lattices, such as direct
imaging~\cite{Campbell, Bakr}, light scattering~\cite{Landig}, and the detection of discrete energy signatures for each
occupation number $n$~\cite{Will, Chen}. Despite these experimental successes, little work has been done to investigate
how these measured distributions depend on parameters of quantum quenches, for example the quench speed. Further
exploration of number distributions and their manipulation is warranted to potentially allow for the optimization of
desirable qualities.
\begin{figure}[b]
\includegraphics[width=85mm]{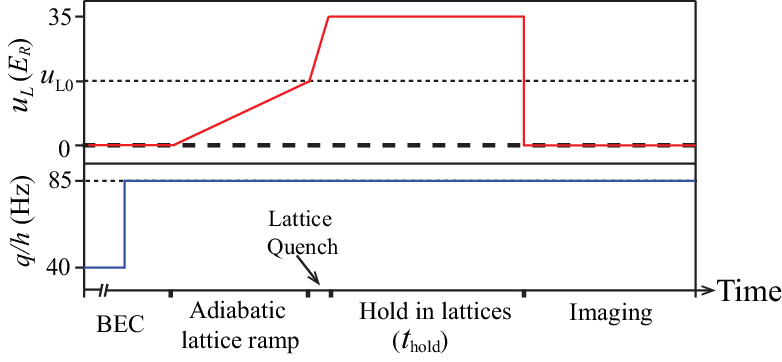}
\caption{Schematic of an experimental cycle of the Quench-\emph{L} sequence (see text). $u_{L0}$ is the intermediate
lattice depth. Axes are not to scale.} \label{sequence}
\end{figure}

With spin-dependent interactions $U_2$, spinor gases exhibit both superfluidity and magnetism while presenting some other
features which are absent in scalar bosons. One notable feature is coherent spin interconversions among multiple spin
components resulting from the competition of $U_2$ and the quadratic Zeeman energy $q$~\cite{Zhao,zhaoUwave,Jiang,Austin}.
By analyzing these spin-mixing dynamics, the signatures of discrete energy levels can be extracted from a Fourier
analysis, and many interesting features of many-body ground states in spinor gases can be revealed~\cite{Mahmud,Chen}.
This approach has been developed in theory by Ref.~\cite{Mahmud} and experimentally verified for adiabatic ramps by our
previous work~\cite{Chen}. In this paper, we extend this method to probe atom number distributions after nonadiabatic
quantum quenches. Our data demonstrate that atom distributions can be manipulated by properly designing the quantum quench
sequence, which may have important applications in attaining different many-body quantum phases. Our observations also
suggest methods of maximizing the fraction of atoms in even Mott lobes which, among other things, may enable future works
to optimize the production of massively entangled states in cold atoms. Another notable aspect of our experiment is the
toolbox it provides for probing spinor atoms at the on-site level, which can be utilized to deduce the coefficients of the
on-site wavefunction and calculate number fluctuations, spin-singlet order parameter, and various other entanglement
observables in homogeneous systems~\cite{Mahmud,Zhao2}.
\begin{figure*}[tb]
\includegraphics[width=170mm]{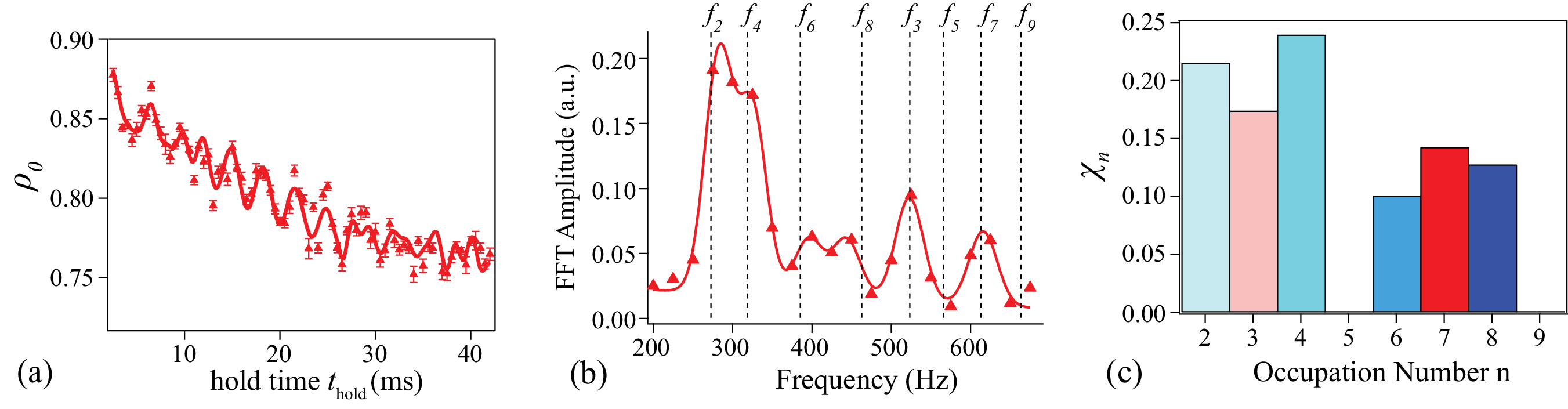}
\caption{(a) Observed dynamics of spin-0 atoms after a Quench-\emph{L} sequence at $v_{\rm{ramp}}=14(1) E_R/\rm{ms}$ and
$u_{L0}=0E_R$ (see text and Fig.~\ref{sequence}). Markers represent the average of approximately 15 repeated shots at the
same conditions and error bars are one standard error. The solid line is a fit based on our empirical model to guide the
eye~\cite{Chen}. (b) Markers show fast Fourier transformations (FFT) over the first 40~ms of $t_{\rm{hold}}$ on the data
shown in Panel~(a). Vertical lines mark the predicted energy signature $f_n$ for each $n$ (see text). The solid line
represents Gaussian fits to each observed peak. (c) Number distributions $\chi_n$ extracted from the FFT spectrum shown in
Panel~(b) (see text).} \label{FFT}
\end{figure*}

Similar to our previous work~\cite{Austin}, we apply the Bose-Hubbard (BH) model and the Gutzwiller approximation to
understand the static and dynamic properties of lattice-trapped spin-1 bosons and express the spin-1 BH Hamiltonian as
\begin{align}
\hat{H} =& -J\sum_{<i,j>,m_F}(\hat{a}_{i,m_F}^\dagger \hat{a}_{j,m_F}+\hat{a}_{j,m_F}^\dagger \hat{a}_{i,m_F})-\sum_{i}\mu_{i} \hat{n}_{i} \notag\\
&+\frac{U_{0}}{2}\sum_{i}\hat{n}_{i} (\hat{n}_{i}-1)+ \frac{U_{2}}{2}\sum_{i}(\vec{F}_{i}^{2} -2 \hat{n}_{i}) \notag\\
&+\frac{1}{2}V_{T}\sum_{i}(i-\frac{L}{2})^2\hat{n}_{i}+q\sum_{i}(\hat{n}_{i,1}+\hat{n}_{i,-1})~. \label{FullHamiltonian}
\end{align}
Here $U_{0}$ is the spin-independent interaction, $J$ is the tunnelling energy, $\mu$ is the chemical potential, $V_T$ is
the external trapping potential, $\vec{F}$ is the spin operator, $\hat{n_i}=\sum_{m_F} \hat{n}_{i,m_F}=\sum_{m_F}
\hat{a}^\dagger_{i,m_F} \hat{a}_{i,m_F}$ is the atom number operator at site-$i$, $\hat{a}_{m_F}$
($\hat{a}^\dagger_{m_F}$) is the boson destruction (creation) operator for the spin-$m_F$ component, $m_F$ can be 0 or
$\pm1$ for $F=1$ atoms, and $L$ is the number of lattice sites~\cite{Austin,Mahmud}. In deep lattices where tunnellings
among neighboring lattice sites are negligible, Eq.~\eqref{FullHamiltonian} can be further decoupled into the single site
Hamiltonian~\cite{Mahmud,Chen},
\begin{equation}
\hat{H} = \frac{U_{0}}{2} \hat{n} (\hat{n}-1) + \frac{U_{2}}{2} (\vec{F}^{2} -2 \hat{n})+ q (\hat{n}_{1}+\hat{n}_{-1})
-\mu \hat{n}~, \label{Hamiltonian}
\end{equation}
where $\hat{n} =\sum_{m_{F}}\hat{n}_{m_F}$ is the atom number operator for each lattice site.

In the Gutzwiller approximation, the many-body wave function of the full lattice can be written as a product of single
site states, which for a homogeneous system is $|\phi \rangle = \sum_{n_{1},n_{0},n_{-1}}
C_{n_{1},n_{0},n_{-1}}|n_{1},n_{0},n_{-1} \rangle~$ in the Fock state basis $|n_{1},n_{0},n_{-1} \rangle$. Fock state
coefficients $P(n_1,n_0,n_{-1})=|C_{n_{1},n_{0},n_{-1}}|^2$ define the Fock state number distributions. The Fock state
coefficient for the spin-0 component is then denoted as $\chi_n=\sum_{n_{1},n_{-1}}|C_{n_{1},n_{0},n_{-1}}|^2$, which can
be obtained by performing Fourier analysis of the spin-mixing dynamics~\cite{Mahmud,Chen}. A related quantity is the
number density in each lattice site which is defined as $N(n_0)= \sum_{n_{1},n_{-1}} n_0 |C_{n_{1},n_{0},n_{-1}}|^2$. The
Fock state coefficients and number densities for other spin components can be found by the same method.

For a homogeneous system, $\chi_n$ reveals the on-site atom number statistics which displays a Poissonian behavior in a
superfluid (SF) state and gets number squeezed in the MI regime, while $N(n_0)$ gives the expectation value of the atom
number or the average density of the atoms~\cite{Sansone}. In our experimental systems, however, atoms in cubic lattices
are externally confined in a harmonic trapping potential. This results in an inhomogeneous density profile with different
atom number statistics at individual lattice sites. In the MI state, this leads to a wedding cake structure in the density
profile with constant integer density Mott plateaus. For such an inhomogeneous system, each lattice site therefore has
different on-site number statistics. The many-body wavefunction at site-$i$ can be expressed as $|\phi_i \rangle =
\sum_{n_{i,1},n_{i,0},n_{i,-1}} C_{n_{i,1},n_{i,0},n_{i,-1}}|n_{i,1},n_{i,0},n_{i,-1} \rangle$. The atom number
distributions and number density distributions in our system become site-dependent, e.g., $\chi_{n_{i}}$ and $N(n_{i,0})$
represent the distributions of spin-0 atoms in the \emph{i}-th lattice site. In our experiments, we collect data after
releasing atoms from all trapping potentials, each observed number distribution is thus summed over all lattice sites,
i.e., $\chi_n=\sum_i \chi_{n_{i}}$.

\begin{figure*}[tb]
\includegraphics[width=170mm]{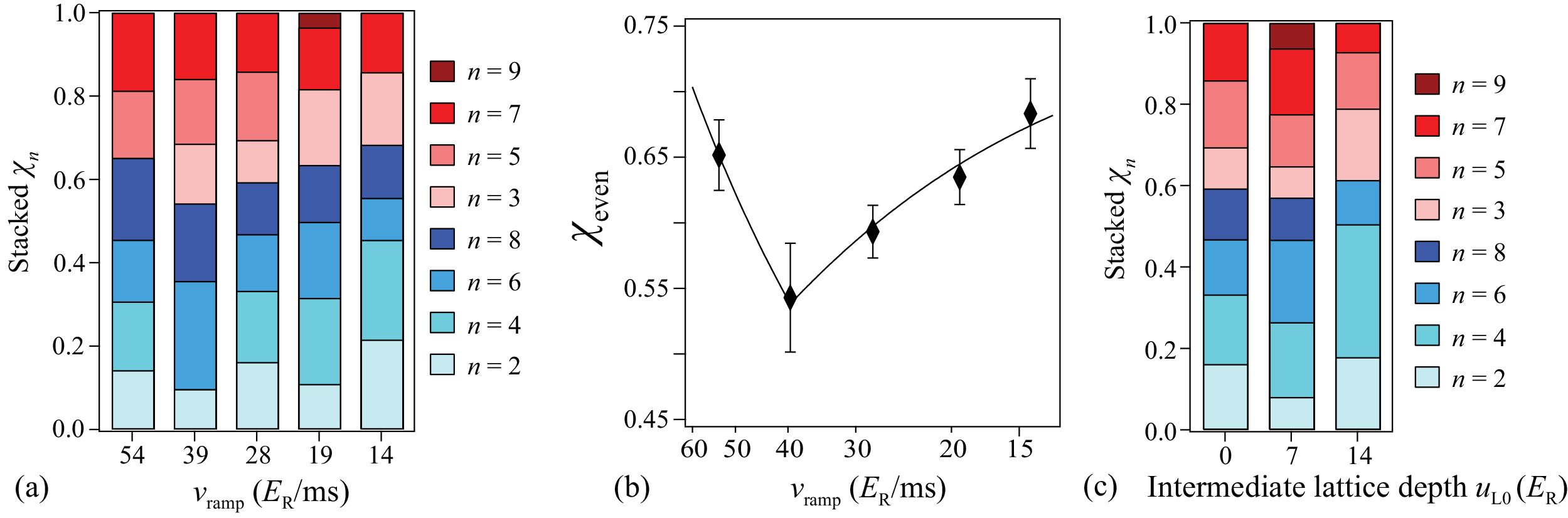}
\caption{(a) Observed number distributions $\chi_n$ after Quench-\emph{L} sequences at various $v_{\rm{ramp}}$ and
$u_{L0}=0E_R$. Shades of blue (red) represent even (odd) occupation numbers $n$ with the shades getting darker as $n$
increases from 2 to 8 (from 3 to 9). The height of each shaded box represents $\chi_n$ for a given $n$, while the combined
height of the blue (red) boxes corresponds to the total number distribution in even (odd) Mott lobes. (b) Diamonds
represent the experimentally found fraction of even Mott lobes $\chi_{\rm{even}}$ for each quench sequence shown in
Panel~(a). The solid line is a linear (an exponential) fit to the data when $v_{\rm{ramp}}$ is faster (slower) than
$39(1)E_R/\rm{ms}$. (c) Similar to Panel~(a) but uses Quench-\emph{L} sequences with various $u_{L0}$ while keeping
$v_{\rm{ramp}}$ at $28(1) E_R/\rm{ms}$ during the lattice ramp from $u_{L0}$ to $35E_R$ (see text).}\label{ExpN}
\end{figure*}

We start each experiment cycle with a spinor BEC of approximately $1.2\times10^5$ sodium ($^{23}$Na) atoms at its SF
ground state, the longitudinal polar state with $\rho_0=1$ and $m=0$. Here $\rho_{m_F}$ is the fractional population of
the $m_F$ state and $m=\rho_{+1}-\rho_{-1}$ is the magnetization. The spinor gases studied in this paper exhibit
antiferromagnetic characteristics because $U_2=0.035U_0>0$~\cite{Chen,Austin}. We then load atoms into a cubic optical
lattice with a lattice spacing of 0.532 microns and continuously quench the lattice depth $u_L$ through the SF to MI phase
transition points via Quench-\emph{L} sequences at different speeds. The lattice depth is in the unit of the recoil energy
$E_R$~\cite{Zhao, Chen, Austin}. Figure~\ref{sequence} shows the schematic of Quench-\emph{L} sequences, in which we first
adiabatically raise $u_L$ to an intermediate value $u_{L0}$ at the ramp speed of $1.4E_R/\rm{ms}$ and then
nonadiabatically quench $u_L$ to a deep final lattice depth $u^{\rm final}_{L}$ at a faster speed $v_{\rm{ramp}}$ to
initiate spin-mixing dynamics. Because $u^{\rm final}_{L}$ is much deeper than the SF-MI transition points, which are
between $16E_R$ and $23E_R$ for our system of $n_{\rm{peak}}=7$, atoms are localized into individual lattice sites by the
end of the quench sequence~\cite{Jiang}. Here $n_{\rm{peak}}$ is the peak occupation number per lattice site in
equilibrium MI states~\cite{Zhao, Chen, Austin}. After the lattice quench, we hold the atoms at $u^{\rm final}_{L}$ for a
holding time $t_{\rm{hold}}$ before abruptly releasing them, and then detect different spin components via a two-stage
microwave imaging process~\cite{Zhao2, Austin}.

While $q/h \lesssim 100$~Hz, spin dynamics similar to those presented in Fig.~\ref{FFT}(a) are observed, where $q$ is the
quadratic Zeeman energy and $h$ is Planck's constant. Consisting of multiple Rabi-type oscillations of frequencies $f_n$,
these dynamics offer an ideal platform to probe the atom number distributions of spinor gases~\cite{Chen, Mahmud}. Here
$f_n=E_n/h$, while $E_n$ is the energy gap between the first excited state and the ground state for a fixed
$n$~\cite{Mahmud,Chen}. When $q$ and $u^{\rm final}_{L}$ are carefully chosen to be large enough that the frequencies of
spin-mixing oscillations at individual $n$ are well separated but $q$ is small enough that the system still displays spin
oscillations after quantum quenches, the resulting spin dynamics allow us to extract the number distributions of our
system. In this paper, all data are collected at $u^{\rm final}_{L}=35E_R$ and $q/h=85$~Hz which offers a good balance
between these conditions. By conducting a fast Fourier transformation (FFT) over the first $40\rm{ms}$ of each observed
spin dynamics, we can precisely determine the spectral weight of each oscillation and thus extract the atom number
distributions after quenches at various $v_{\rm{ramp}}$. We choose 40~ms for each FFT as it is sufficiently long to
clearly resolve each peak while remaining short enough to avoid the number distribution changing significantly with
$t_{\rm{hold}}$ during the data taking. The observed FFT spectra, similar to the one shown in Fig.~\ref{FFT}(b), are then
fit with an eight-Gaussian fitting. By integrating over each peak and dividing by the theoretical spin oscillation
amplitude at each $n$, we can obtain the spectral contributions of individual $n$~\cite{Chen}. As shown in
Fig.~\ref{FFT}(c), when these values are normalized this process gives us the atom number distributions $\chi_n$ in deep
lattices for each integer $n$. Because no spin oscillations occur when $n=1$, all $\chi_n$ shown in this paper reflect the
normalized number distributions after the $n=1$ Mott lobe is excluded.

\begin{figure}[tb]
\includegraphics[width=85mm]{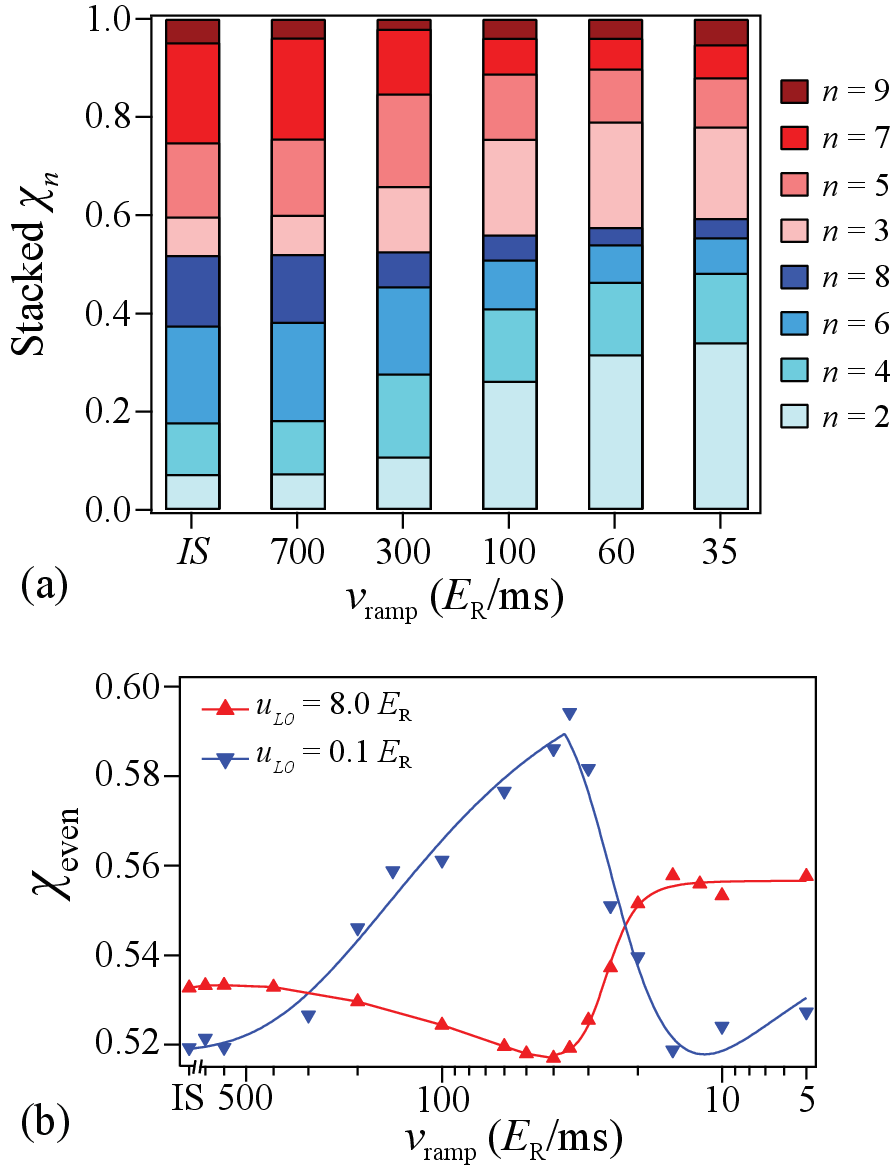}
\caption{(a) Predicted number distributions $\chi_n$ derived from our numerical simulations for 2D lattice-confined sodium
spinor gases after Quench-\emph{L} sequences with $u_{L0}=0.1E_R$ at various quench speeds $v_{\rm{ramp}}$ and for $IS$,
the initial superfluid ground state. Shades of blue (red) represent even (odd) $n$ with the shades getting darker as $n$
increases from 2 to 8 (3 to 9). The height of each shaded box represents the predicted $\chi_n$ for a given $n$, while the
combined height of the blue (red) boxes represents the total number distributions in even (odd) Mott lobes. (b) Markers
represent fractions of even Mott lobes extracted from simulated lattice quenches similar to those shown in Panel~(a) but
at various $v_{\rm{ramp}}$ and two $u_{L0}$. Solid lines are fitting curves to guide the eye.} \label{theoryRamp}
\end{figure}
The observed atom number distributions after Quench-\emph{L} sequences at various $v_{\rm{ramp}}$ are displayed in
Fig.~\ref{ExpN}(a). Because the realizations and manipulations of some important quantum states (e.g. spin singlets) of
ultracold atoms depend on the increased presence of even Mott lobes~\cite{Jiang,Demler,Imambekov, Lacki,Snoek,Yip}, one
parameter of particular significance that we wish to probe is $\chi_{\rm{even}}$, the fraction of atoms in even Mott
lobes. Figure~\ref{ExpN}(b) shows the observed $\chi_{\rm{even}}$ as a function of the lattice quench speed. As seen in
Fig.~\ref{ExpN}(b), the distributions found at the fastest tested speed of $v_{\rm{ramp}}=54(1) E_R/\rm{ms}$ have a
relatively high $\chi_{\rm{even}}$, and there is a clear dip in $\chi_{\rm{even}}$ as the quench speed is lowered from
$54(1) E_R/\rm{ms}$ to $39(1) E_R/\rm{ms}$, which then increases exponentially with $v_{\rm{ramp}}$ back to a relatively
high $\chi_{\rm{even}}$. One notable result is that most observed $\chi_{\rm{even}}$ shown in Fig.~\ref{ExpN}(b) are
larger than the predicted $\chi_{\rm{even}}$ of $0.55$ for equilibrium MI states at $n_{\rm{peak}}=7$. We find no
discernible FFT spectra can be obtained after Quench-\emph{L} sequences at slow speeds of $v_{\rm{ramp}}<10E_R/\rm{ms}$
because the amplitude of spin-mixing oscillations after the quenches appears to rapidly diminish as $v_{\rm{ramp}}$
decreases. As elaborated in our previous work~\cite{Chen}, to study number distributions after adiabatic ramps, we thus
have to use a Quench-\emph{Q} sequence, in which atoms first adiabatically cross the SF-MI phase transitions in a high
magnetic field of $q\gg U_2$ to ensure the atoms remain in their ground states and then spin dynamics are initiated by
quenching the magnetic field to a desired final $q$.

We also examine the effects of varying the intermediate lattice depth $u_{L0}$ at the start of the quench in
Quench-\emph{L} sequences in Fig.~\ref{ExpN}(c). In these experiments, the lattice ramp speed $v_{\rm{ramp}}$ is kept at
$28(1) E_R/\rm{ms}$ during the quenches from various $u_{L0}$ to the final lattice depth of $35E_R$. Our data in
Fig.~\ref{ExpN}(c) indicate that the observed $\chi_{\rm{even}}$ only weakly depends on $u_{L0}$ at this $v_{\rm{ramp}}$.
As $u_{L0}$ increases and approaches SF-MI transition points, the maximum $n$ extracted from the experimental FFT spectra
appear to decrease and get close to the predicted $n_{\rm{peak}}$ for equilibrium MI states, which confirms lattice ramps
become more adiabatic at larger $u_{L0}$.

Our experiments are performed in 3D inhomogeneous lattice-confined spin-1 spinor gases. An exact many-body simulation of
such systems has not yet been reported in the literature either for equilibrium states or nonequilibrium
dynamics~\cite{Austin}. This numerical problem is prohibitively difficult due to high Hilbert space dimensions of on-site
spin-1 atoms and the inhomogeneous nature of the systems, and hence feasible theoretical simulations are limited to one
and two dimensions (2D). Figure~\ref{theoryRamp}(a) shows typical simulation results of atom number distributions
performed in systems similar to our experimental system but in 2D after Quench-\emph{L} sequences at various quench speeds
and also for the initial SF ground state which corresponds to an infinitely fast ramp in Quench-\emph{L} sequences. Each
predicted distribution shown in Fig.~\ref{theoryRamp} is averaged over all lattice sites and only includes the
experimentally observable Mott lobes of $n\geq2$.

We find qualitative agreements between our 2D theoretical simulations and 3D experimental results, despite quantitative
disparities which are expected due to the difference in dimensionality. For example, similar to our experimental data
shown in Fig.~\ref{ExpN}(a), the simulated atom number distributions in Fig.~\ref{theoryRamp}(a) indicate a dependence on
the lattice quench speed $v_{\rm{ramp}}$. The predicted even fractions after simulated quenches at various $v_{\rm{ramp}}$
(see Fig.~\ref{theoryRamp}(b)) also display some similarities to our experimental observations illustrated in
Fig.~\ref{ExpN}(b), e.g., $\chi_{\rm even}$ exponentially increases from a value around the predicted $\chi_{\rm even}$
for the initial SF ground state as $v_{\rm{ramp}}$ is lowered, and $\chi_{\rm even}$ reaches its maximum at a nonadiabatic
quench speed. Our theoretical results indicate that $\chi_{\rm{even}}$ should decrease to the value for equilibrium MI
states as $v_{\rm{ramp}}$ continues to decrease but because we are unable to observe spin oscillations at very small
$v_{\rm{ramp}}$, we have not been able to observe this experimentally. Differences between the two theory curves in
Fig.~\ref{theoryRamp}(b) also imply that it is possible to optimize the even fraction $\chi_{\rm even}$ by properly
designing the quantum quench sequence, e.g., larger maximum achievable $\chi_{\rm even}$ may be realized at smaller
$u_{L0}$. These results suggest that atoms go through complex spatial dynamics while redistributing within the harmonic
trap during the quantum quenches, and number distributions reach an equilibrium value when the quench speed is
sufficiently slow to ensure the atoms initially located in the trap center have enough time to move towards the trap
boundaries and equilibrate. Note that Gutzwiller approximation ignores all inter-site correlations and entanglement to
compute the ground state properties and the dynamics. The spinor system near SF-MI transitions, however, is a highly
correlated system where inter-site correlations may need to be considered for a more rigorous analysis. Other numerical
methods may be better suited to study this phenomenon, but we are further limited by dimensionality, i.e., we can only
conduct simulations in 1D or 2D systems rather than 3D systems. Density Matrix Renormalization Groups (DMRG) algorithms in
particular might yield valuable insight but are only efficient in 1D systems~\cite{Austin}. An efficient high performance
computation of 3D lattice-confined spinor systems is a planned future research avenue for improving the theory-experiment
comparisons.

In conclusion, we have experimentally demonstrated the manipulation of atom number distributions of lattice-confined
spinor gases after nonadiabatic quantum quenches. Our data have illustrated methods of maximizing the fraction of atoms in
even Mott lobes, which have applications in attaining various quantum magnetic phases including massively entangled
states. Qualitative agreements have also been found between our experimental data and numerical simulations using
time-dependent Gutzwiller approximations in two-dimensional systems.

\begin{acknowledgments}
We thank the National Science Foundation and the Noble Foundation for financial support.
\end{acknowledgments}

\end{document}